\documentclass[prl,aps,twocolumn,groupedaddress,showpacs,floatfix]{revtex4}
\usepackage{amsmath,amssymb,multirow,epsfig,bm}

\begin{document}

\title{Heavy-Light Few Fermion Clusters at Unitarity}

\author{S. Gandolfi and J. Carlson}
\affiliation{Theoretical Division\\
Los Alamos National Laboratory \\
Los Alamos, NM 87545 }

\begin{abstract}
We examine the physics of two, three, and four heavy fermions interacting with a single light fermion via short-range interactions. Four-particle bosonic 
Efimov states have proven important experimentally and 
also been the subject of significant theoretical effort.  
Similar fermionic systems 
are just now being investigated. We find that with some simple interactions 
the four- and five-particle  states collapse to the interaction range 
at smaller mass ratios than the three-body state, and also before
larger clusters can collapse. 
These  states and their excitations can be studied in cold atom experiments,
providing unique insights into the role of few-body systems in 
many-body physics.
\end{abstract}

\date{\today}
\maketitle

  The few- and many-body physics of heterogeneous mixtures of cold Fermi atoms 
is of great interest both experimentally\cite{Wille:2008,Taglieber:2008} 
and theoretically\cite{Liu:2003,Braaten:2006,
Wu:2006,Stecher:2007,Combescot:2007,
Baranov:2008,Nishida:2008,Petrov:2005,Guo:2008}.  
For many-body systems, different mass ratios may allow for `exotic' superfluid
phases,  or phases of mixtures of different condensates to be observed experimentally. Few-body mixtures of heavy and light fermions also allow us
to study the role of new classes of fermionic few-body resonances, 
in particular searching for regimes in which four- and five-particle states
can play a significant role.

The impact of few-body resonances in cold atomic gases has been confirmed recently for bosons, where Efimov states have been predicted theoretically and found experimentally\cite{Hammer:2007,vonStecher:2009,Pollack:2009,Dincao:2009,Mehta:2009}.  These states can have significant effects on 
loss rates in cold-atom experiments, and are of course very intriguing in their own right.  In the present work, we examine potential few-body fermion states and their impact.  Such states are clearly very different from the bosonic Efimov states, but offer a potentially intriguing prospect for future studies.

In the many-body heterogeneous system, the simplest phase diagram 
at unitarity for a mass ratio  relevant to mixtures of K and Li has been 
examined in Ref. \citenum{Gezerlis:2009}.  A normal state is found 
at large polarization, particularly where the number of heavy fermions greatly 
exceed the number of light fermions.  The resulting energy is very small, though, suggesting that  intriguing states perhaps including multi-particle condensates may arise near this regime.

Theoretical investigations of the few-body problem with multiple heavy fermions and a one light fermion have already provoked a great deal of interest.  
Interestingly, calculations
for two heavy and one light fermions find regions where collapse is possible at mass ratios significantly below the 
critical value of $~$13.6 where collapse and associated Efimov states are universal.\cite{Nishida:2008}  The few-body interactions must be fine-tuned to produce these states, and hence the few-body physics is non-universal.
In this study, Nishida et al., discuss the potential effects of these states
in many-body experiments, finding the potential to conduct examinations of strongly-interacting gases of dimers and trimers.
Our results indicate that an even richer phase structure may be appropriate,
as four- and five-body resonances can appear in addition to the resonances of two heavy and one light particle.  In fact, these resonances can appear at lower mass
ratios than the two-heavy one-light (2H1L) system.

{\it Model Hamiltonian:} We investigate this non-universal few-body regime by calculating ground states 
in a model system of three-, four- and many-fermions interacting with a single light particle via short-range interactions.
The Hamiltonian considered for one light particle of mass $m$ and N heavy fermions of mass $M$ is:
\begin{eqnarray}
H \ & = & \ - \frac{\hbar^2}{2 m} {\nabla_l}^2 + \sum_{h=1}^N 
- \frac{\hbar^2}{2 M} {\nabla_h}^2 + \sum_h V_2 (r_{lh}) \nonumber \\
& & \ \   + \ \sum_{h1<h2} V_3 (r_{lh1},r_{lh2}),
\end{eqnarray}
where 
$l$ labels the light particle, $h$ the heavy particles,
the two-body potential $V_2$ acts only between light and heavy fermions,
and the three-body potential is a function of the separations between the
light and the two heavy particles.  In all calculations the heavy-light effective mass
$\mu = Mm/(M+m)$ is used, along with the effective range of the two-body interaction $r_{eff}$ to set the scales of energy and distance. The two-body interactions are tuned to unitarity or infinite scattering length.  We use both Poschl-Teller and Gaussian two-body potentials:
\begin{eqnarray}
V_2 & = & -2 \frac{\hbar^2 }{ 2 \mu} \lambda^2 \cosh^{-2}( \lambda r) \\ \nonumber
V'_2 & = & -2 v_2^g \frac{\hbar^2 }{  2 \mu} \lambda^2 \exp [ - ( \lambda r)^2/ 2],
\end{eqnarray}
where $v_2^g \approx 0.671$ is adjusted to unitarity . We also include
a three-body interaction with the same range as the two-body interaction
for each pair:
\begin{equation}
V_3 (r,r') =  v_3^0 \ \frac{\hbar^2 }{ 2 \mu} \lambda^2 \cosh^{-2} ( \lambda r) \ \cosh^{-2}( \lambda r').
\end{equation}

    The two-heavy one-light problem has been investigated by Nishida, et al.\cite{Nishida:2008},
who found a regime for $ 8.6 < M/m < 13.6 $ where the interaction can be fine-tuned to produce three-body resonances. In a many-body system these states would yield an interacting gas of dimers and trimers.  In this study they explicitly assume that collapse of four- and five-body states are not favored.   However, for specific sets of geometrically symmetric heavy-particle coordinates (equilateral triangles and regular tetrahedrons, respectively), Nishida\cite{Nishida:2008a} used the Born-Oppenheimer approximation to study the static potential between the heavy fermions.  Significant additional attraction was found for these multi-particle states, suggesting it is at possible that the larger clusters may be bound at lower mass ratios than the two-heavy one-light system.

{\it Methods:} We examine this possibility using Quantum Monte Carlo techniques. The calculations employ variational states to give a variational upper bound to the ground-state energy.  We assume a trial wave function $\Psi_T = \phi_L \Phi_H$, where  $\phi_L$ is a positive definite function of the light particle coordinates, and $\Phi_H$ is an anti-symmetric state of the heavy particle coordinates.  All coordinates are measured from the system center-of-mass to avoid spurious CM motion.

The calculations are variational and hence produce an upper bound to the
ground state of the model Hamiltonian.  We believe that these results are likely to be
accurate in most regimes, as the nodal surfaces $(\Psi_T = 0)$ are quite simple for these systems.  For two, three, or four particles it is possible to put all the fermions in relative p-waves; larger systems require higher partial waves or radial excitations. The simplest possible nodal surfaces are, as in the Born-Oppenheimer approximation, independent of the position of the light particle. They are simply the projected length, projected area, and volume of the line, triangle, and tetrahedron 
connecting the two, three, and four heavy fermions, respectively:
\begin{eqnarray}
\Phi_H^2 & = & ( {\bf r}_{1} - {\bf r}_{2})) \cdot {\hat z}.   \\
\Phi_H^3 & = & {\bf r}_{3,12}  \times ( {\bf r}_{1} - {\bf r}_{2}) \cdot {\hat z}. \nonumber \\ 
\Phi_H^4 & = & {\bf r}_{4,123} \cdot ({\bf r}_{3,12}  \times ( {\bf r}_{1} - {\bf r}_{2})),  \nonumber
\end{eqnarray}
where ${\bf r}_{i,jk}$ is used to indicate the relative coordinate of particle $i$ from
the center of mass of the pair or triplet $jk$.
These ground states have angular momentum $L = 1,1,0$, respectively, for N = 2,3,4.  Slightly lower energies are obtained in the calculations reported here by putting the heavy particles in s- and p-wave orbitals measured from the system CM,
yielding additional variational freedom.

\begin{figure}
\vspace{-0.2in}
\begin{center}
\includegraphics[height=2.3in,angle=0]{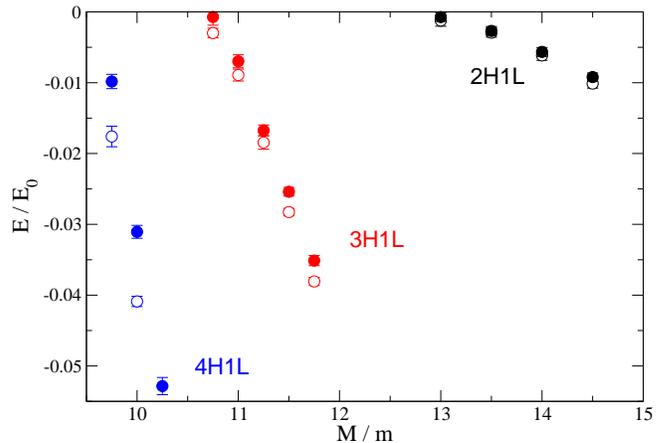}
\caption{(color online) Binding Energies versus mass ratio M/m for 4H1L, 3H1L, and 2H1L systems with attractive two-body potentials.  Energies are given in terms of $E_0$, energies less than zero correspond to systems that will collapse as the range of the interaction goes to zero. Filled symbols correspond to the Poschl-Teller two-body potential and open symbols to the Gaussian potential.} 
\label{fig:twobody}
\end{center}
\vspace{-0.2in}
\end{figure}

{\it Results:} The results for the ground-state energies of the two-, three-, and four- heavy, one light systems are shown in Figure \ref{fig:twobody} for these specific
two-body interactions. All energies are plotted in units of $E_0 \equiv  \hbar^2 / ( 2 \mu r_{eff}^2)$, where $r_{eff}$ is the effective range of the interaction.  For these two-body interactions the regimes where $E \lesssim 0$ are shown, 
nevertheless they would collapse to large negative energies and zero radius as the two-body effective range is reduced. The few-particle system will collapse at this point independent of whether the particles are confined in a trap or not.
 
For these interactions we find that the four heavy one light system collapses at a much smaller mass ratio than the three heavy one light which, in turn, collapses before the two heavy one light. The additional attraction obtained from the
light particle orbiting multiple heavy fermions is sufficient to bind them
quite deeply. The increase in binding with mass ratio is very rapid for the
3H1L and particularly for the 4H1L system. For these simple interactions,
the binding of the 3H1L and 4H1L clusters are very large at the mass ratio where the
2H1L system is near threshold.  Results for the  Poschl-Teller potential
are given in Table \ref{tab:largebinding}.

\begin{table}
\begin{tabular}[c]{|r|r|r|}
\hline
 & M/m = 13 & M/m = 14  \\
 \hline
 2H1L & -0.001(1) & -0.0057(6) \\
 3H1L & -0.088(1) & -0.135(1) \\
 4H1L & -0.340(1) & -0.454(1) \\
 \hline
\end{tabular}
\caption{Binding energies for the 3H1L and 4H1L system at mass
ratios near threshold for the 2H1L system.  Energies are given in terms of $E_0$
for the Poschl-Teller potential.}
\label{tab:largebinding}
\end{table}

It could be possible that adding additional
fermions would bind the system at even smaller mass ratios.  To test this
assertion we have performed calculations of the 9H1L and 10H1L systems using
the same Quantum Monte Carlo techniques. These particle numbers are suggested by filling single-particle states in the d (L=2) wave and a radial excitation
into the 2s state. Adding additional fermions does not seem to reduce the minimum mass ratio further, presumably the additional energy cost of putting heavy fermions in higher angular-momentum or higher-energy radial states is too large to bind the overall system.  There exists a mass ratio where a single light particle 
could bind an infinite number of heavy particles, however in our calculations to date this ratio appears significantly larger than the ratio where smaller systems
can collapse.

We have calculated the root-mean-square radius for the heavy and light particles
for the clusters near threshhold. The results for a few cases are summarized
in Table \ref{tab:radii}.  For the 2H1L system we find that the light-particle
rms radius at very weak binding is about 3-4 times the heavy particle radius, consistent with simple expectations.  The difference in radii decreases as we go to the larger systems and larger binding, though. 
In the table we report VMC results for the trial wave function
$\Psi_T$ and the QMC results.  The large extrapolations make very
accurate results difficult to extract, but the trend is clear.

\begin{table}
\begin{tabular}[c]{|r|r|r|r|r|r|}
\hline
        & M/m & VMC L & VMC H & QMC L & QMC H \\
\hline
2H1L & 14.5 & 0.78 & 0.18 & 1.14 & 0.30 \\
3H1L & 11.75 & 0.30 & 0.18 & 0.42 & 0.24 \\
4H1L & 10.25 & 0.30 & 0.24 & 0.30 & 0.24 \\
 \hline
\end{tabular}
\caption{RMS radii for the heavy and light particles for the clusters.
VMC and extrapolated QMC result are given, results are for the 
Poschl-Teller potential and all radii are given in units  
of $r_{eff}$ }
\label{tab:radii}
\end{table}

We've also examined the single-particle densities of the clusters.
Because of the nodal structures defined above, the heavy particles have
very small density near the cluster's center-of-mass. In the case of the simple
trial functions above, the density is exactly zero at the origin. The light
particle, in contrast, has a maximum density at the origin, preferring to
occupy a space in the middle of the heavy fermions.

\begin{figure}
\vspace{-0.2in}
\begin{center}
\includegraphics[height=2.5in,angle=0]{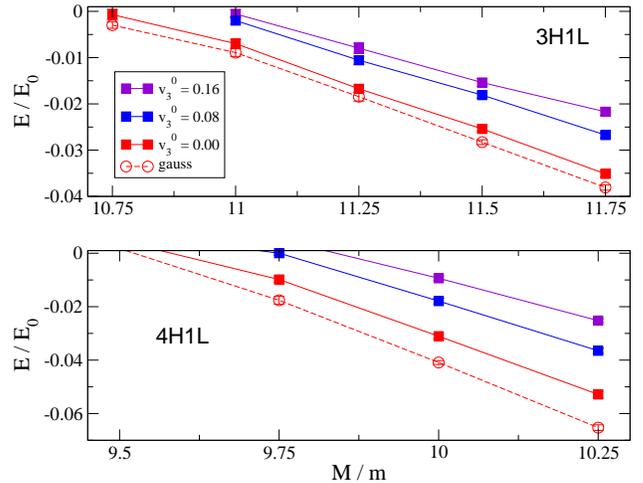}
\caption{(color online) The effect of three-body interactions on the binding energies of 3H1L. The three-body interaction can shift the mass threshhold for collapse significantly.}
\label{fig:threebody}
\end{center}
\vspace{-0.2in}
\end{figure}

It is clear from our results that different two-body potentials with the same effective range give different binding energies. In fact the mass ratio where the curves intercept $E=0$ changes with different two-body interactions.  To further quantify these results, we've also studied these systems including a repulsive three-body interaction $V_3$.  The three-body interaction also shifts the  zero binding energy intercept.  These results underscore the non-universal nature of this regime.
As an example, in  Figure \ref{fig:threebody} we compare results with- and without
a three-body interaction near threshhold for the  3H1L and 4H1L systems.  Even relatively small amplitudes of the three-body interaction shift the
curves significantly, as is to be expected for a non-universal system. 
The 2H1L systems behaves similarly, with a significant impact of the three-body force.  All systems can become unbound for a repulsive three-body
interaction with a strength $V_3$ of order unity.

{\it Future Possibilities:}
It would be interesting to know if a single three-body interaction parameter
can simultaneously describe the binding of the 2H1L, 3H1L, and 4H1L
systems.  This has been found to be the case in few-body bosonic systems, or in
fermionic systems where the number of spin degrees of freedom is larger
than the number of particles.  In particular this relation has been used to explain
the relative binding of the deuteron, 3H, and 4He nuclei.   Of course the
interaction here cannot be a pure contact interaction but would require
momentum-dependent terms for the two heavy
fermions, which must be in an odd relative partial wave, coupled to one boson.

We are presently investigating the possibility of bound excited states in these
systems in certain ratios of mass regimes,
perhaps related by simple scaling laws.  The large binding of the heavier systems relative to the smaller gives some reason to believe this may be possible.
One can imagine states similar to those found for bosons where one heavy
particle is in an excited loosely bound state.  It would have to be in an s-wave
or other low partial wave to have a significant attraction to the remaining N-1
particle cluster.
Finally, it could be extremely interesting to study the systems described here
in the many-body context.  Optical lattices can be used to set the
effective mass ratio as described in \cite{Nishida:2008}.  One could in principle set the mass ratio and the relative population of the heavy and light species
 to obtain strongly-interacting gases of 4H1L and 3H1L clusters.  Many other possibilities can also be imagined. 
 
 {\it Acknowledgements:}  The authors would like to thank D. Son, Shina
 Tan, H.-W. Hammer, D. Blume, and K. E. Schmidt for interesting discussions. The work of S.G. and J.C. is supported by the Nuclear Physics Office
 of the US Department of Energy, computations were carried out through Open Supercomputing at LANL and at NERSC.
 
During preparation of this manuscript we became aware of related independent
work by D. Blume and K. M. Daily.  Preliminary initial discussions of this physics
occured at the Institute for Nuclear Theory at the University of Washington in
March 2010.\cite{Carlson:2010}


\begin{thebibliography}{99}

\bibitem{Taglieber:2008} M. Taglieber, A.C. Voigt, T. Aoki, T.W. Hansch, K. Dieckmann, Phys. Rev. Lett. {\bf 100}, 010401 (2008).

\bibitem{Wille:2008}  E. Wille {\it et al.}, Phys. Rev. Lett. {\bf 100}, 053201 (2008).


\bibitem{Liu:2003} W. V. Liu and F. Wilczek, Phys. Rev. Lett. {\bf 90} 047002 (2003).

\bibitem{Braaten:2006} E. Braaten and H.-W. Hammer, Phys. Rept. {\bf 428} 259 (2006).

\bibitem{Wu:2006} S.T. Wu, C.-H. Pao, and S.-K. Yip, Phys. Rev. B {\bf 74}, 224504 (2006)

\bibitem{Stecher:2007} J. von Stecher, C. H. Greene, and D. Blume, Phys. Rev. A {\bf 76}, 053613 (2007).

\bibitem{Combescot:2007} R. Combescot, A. Recati, C. Lobo, F. Chevy, Phys. Rev. Lett. {\bf 98}, 180402 (2007).

\bibitem{Baranov:2008} M. A. Baranov, C. Lobo, and G. V. Shlyapnikov, Phys. Rev. A {\bf 78}, 033620 (2008).

\bibitem{Nishida:2008} Yusuke Nishida, Dam Thanh Son, and Shina Tan, Phys. Rev. Lett. {\bf 100}, 090405 (2008).

\bibitem{Petrov:2005} D.S. Petrov, C. Salomon, G.V. Shlyapnikov, J. Phys. B: At. Mol. Opt. Phys. 
{\bf 38}, S645 (2005).

\bibitem{Guo:2008} H. Guo {\it et al.}, arXiv:0812.3121 (2008).

\bibitem{Hammer:2007} H.-W. Hammer and L. Platter, Eur. Phys. J. A {\bf 32}, 113 (2007).

\bibitem{vonStecher:2009} J. von Stecher, J. P. D'Incao, and Chris H. Greene,
Nature Physics {\bf 5}, 419 (2009).

\bibitem{Pollack:2009} S. E. Pollack, D. Dries, and R. G. Hulet, Science {\bf 326}, 1683 (2009).

\bibitem{Dincao:2009} J. P. D'Incao, J. von Stecher, and Chris H. Greene,
Phys. Rev. Lett {\bf 103}, 033004 (2009).

\bibitem{Mehta:2009} N. P. Mehta, S. T. Rittenhouse, J. P. D'Incao, J. von Stecher, C.H. Greene, Phys. Rev. Lett. {\bf 103}, 153201 (2009).

\bibitem{Gezerlis:2009} A. Gezerlis, S. Gandolfi, K. E. Schmidt, and J. Carlson,
Phys. Rev. Lett. {\bf 103}, 060403 (2009).


\bibitem{Nishida:2008a} Yusuke Nishida, Phys. Rev. A {\bf 79}, 013629 (2009).

\bibitem{Carlson:2010} see talks at {\it \\
http://www.int.washington.edu/talks/WorkShops/int\_10\_46W}.

        
\end{thebibliography}
\end{document}